\documentclass[12pt]{iopart}
\pdfoutput=1
\usepackage{natbib}

\usepackage[pdftex]{graphicx}
\usepackage[english]{babel}
\usepackage[pdftex,usenames]{color}
\usepackage{hyperref}
\usepackage{amsfonts}

\newcommand{\av}[1]{\langle #1 \rangle}

\newcommand{\be}{\begin{equation}}
\newcommand{\ee}{\end{equation}}
\newcommand{\bea}{\begin{eqnarray}}
\newcommand{\eea}{\end{eqnarray}}

\newcommand{\conf}{\vec{n}}
\newcommand{\confx}{\vec{x} }
\newcommand{\dd}{\partial }

\begin{document}

\title{Ordering dynamics of the multi-state voter model}

\author{Michele Starnini$^1$, Andrea Baronchelli$^2$, and Romualdo
  Pastor-Satorras$^1$} 

\address{$^1$Departament de F\'\i sica i Enginyeria Nuclear,
  Universitat Polit\`ecnica de Catalunya, Campus Nord B4, 08034
  Barcelona, Spain}

\address{$^2$Laboratory for the Modeling of Biological and
  Socio-technical Systems, Northeastern University, Boston MA 02115
  USA}

\begin{abstract}
  The voter model is a paradigm of ordering dynamics. At each time
  step, a random node is selected and copies the state of one of its
  neighbors. Traditionally, this state has been considered as a binary
  variable. Here, we address the case in which the number of states is
  a parameter that can assume any value, from $2$ to $\infty$, in the
  thermodynamic limit. We derive mean-field analytical expressions for
  the exit probability, the consensus time, and the number of
  different states as a function of time for the case of an arbitrary
  number of states. We finally perform a numerical study of the model
  in low dimensional lattices, comparing the case of multiple states
  with the usual binary voter model. Our work sheds light on the role
  of the parameter accounting for the number of states.
\end{abstract}


\maketitle

\section{Introduction}
\label{sec:introduction}

Models of ordering dynamics have since long been considered as
paradigms of opinion dynamics and consensus formation in social
systems \cite{Castellano09}.  Most of them share the
fundamental feature that order results from the self-organization of
local and usually short-range pairwise interactions between agents, as 
it is well illustrated by the simplest and most analyzed of
them, the so-called voter model \cite{Clifford73,Holley:1975fk}. In
its basic formulation, the voter model is defined as follows: Each
individual in a population (agent) is endowed with a binary spin
variable, representing two alternative opinions, and taking values
$\sigma=\pm1$. At each time step, an agent $i$ is selected at random 
together with one nearest neighbor $j$  and the state of the
system is updated as $\sigma_i := \sigma_j$, the first agent copying
the opinion of its neighbor.  Starting from a disordered initial
state, this dynamics leads in finite systems to a uniform state with
all individuals sharing the same opinion (the so-called consensus).

% The interest of the voter model, apart from its
% relation to social science modeling, lies in the fact that it is one
% of the few non-equilibrium stochastic models that can be exactly
% solved in any number of dimensions
% \cite{liggett99:_stoch_inter,KineticViewRedner}.

From the point of view of social dynamics, the interest in this kind of
models is mainly focused on the way in which consensus is reached. The
approach to this state is characterized in terms of the exit
probability $E(x)$ and the consensus time $T_N(x)$, defined as the
probability that the final state corresponds to all agents in the
state $+1$ and the average time needed to reach consensus in a system
of size $N$, respectively, when starting from a homogeneous initial
condition with a fraction $x$ of agents in state $+1$
\cite{Castellano09}.  Due to its simplicity, the voter model dynamics
can be exactly solved in regular lattices for any number of dimensions
\cite{liggett99:_stoch_inter,KineticViewRedner}. Thus, considering the
average conservation of magnetization $m = \sum_{i=1}^{N} \sigma_i
/N$, it can be shown that the exit probability is always a linear
function, $E(x) = x$. On the other hand, the consensus time starting
from the homogeneous symmetric initial condition $x=1/2$ scales with
system size $N$ as $T_N(1/2) \sim N_\mathrm{eff}$, with
$N_\mathrm{eff} \sim N^2$ in $d=1$, $N_\mathrm{eff} \sim N \log N$ in
$d=2$, and $N_\mathrm{eff} \sim N$ in $d>2$ (at the mean-field level)
\cite{KineticViewRedner}. Finally, the dependence of consensus time
with the initial density of $+1$ spins, starting from homogeneous
initial conditions, takes the form
\begin{equation}
  \label{eq:1}
  T_N(x) = - N_\mathrm{eff} \left[ x \ln (x) + (1-x) \ln (1-x)
  \right], 
\end{equation}
for $d\geq2$ \cite{1751-8121-43-38-385003}. 

Different variants of the voter model have been considered in the
past, including the presence of quenched disorder in the form of
``zealots'' which do not change opinion
\cite{PhysRevLett.91.028701,1742-5468-2007-08-P08029}, memory and
noise reduction \cite{0295-5075-77-6-60005}, inertia
\cite{PhysRevLett.101.018701}, non-conservative voters
\cite{Lambiotte08}, non-linear interactions
\cite{durretnonlinear,Deoliveira93}, etc.; see
Ref.~\cite{Castellano09} for an extended bibliography on this subject.
A variant that has been considered in several contexts is the
\textit{multi-state} voter model, in which each agent can be in one of
$S$ different exclusive states or opinions, in analogy of the Potts
model \cite{Wupottsmodel}. The multi-state voter model has been
considered in the past theoretically in terms of mappings of
coarsening of the Potts model on the Ising model with constant
magnetization \cite{PhysRevE.52.244} or in terms of duality properties
\cite{1742-5468-2008-05-P05006} and has found applications in
understanding the fragmentation transition in adaptive networks
\cite{2012arXiv1201.5198B}, in neutral models of biodiversity
\cite{hubbell2001unified,McKane200467} or in ecological models
\cite{pigolottieco2005}. Variants of the pure multi-state voter model,
introducing non-equivalent states, have also been discussed in the
literature
\cite{0295-5075-85-4-48003,0305-4470-36-3-103,1367-2630-8-12-308}.

In this paper we focus in the study the ordering dynamics of the
symmetric multi-state voter model, focusing in particular on the limit
of a large number of initial states. Each agent can be in one of $S$
different but dynamically equivalent states. Agents follow the same
dynamical update rules than in the binary version, with time being
update at every dynamical step as $t \to t + 1/N$. Expressions for the
consensus time in this model at the mean-field level have already been
provided in the literature
\cite{Tavare1984119,Cox:1989fk,baxter2007}. The derivations presented
so far rely however on heavy mathematics.  Here, building on the
Fokker-Planck formalism presented in
Ref.~\cite{blythe07:_stoch_model}, we rederive in a simple way the
expressions for the exit probability and the consensus times in the
general case of $S$ states.  We find that the consensus time increases
very slowly with the number of states, and its difference with the
binary case saturates as $S \rightarrow \infty$; we rationalize this
finding by comparing it with the case of the mutant invasion in the
ordinary two-state voter model.  We also investigate the decay of the
number of states as a function of time, providing a mean-field
expression in excellent agreement with simulations. We finally
consider the dynamics of the multi-state voter model on low
dimensional lattices. Lacking of specific analytical insights, we
compare the numerically observed phenomenology to the mean-field case,
and point out similarities and differences, focusing on the effect of
the number of initial states and their configuration.

The paper is structured as follows.  Sec.~\ref{sec:multi-state-voter}
is devoted to the analysis of the mean-field multi-state voter
model. Sec.~\ref{sec:finite-dimens-latt} reports on numerical
experiments concerning the low-dimensional case. Finally, 
Sec.~\ref{sec:conclusions} presents our conclusions.

\section{Mean-field analysis}
\label{sec:multi-state-voter}

The form of the consensus time in the multi-state voter model at the
mean-field level has been discussed in the past, mainly in the context
of population genetics dynamics
\cite{Tavare1984119,Cox:1989fk,baxter2007}. Here we present a simple
derivation of this expression, based in the Fokker-Plank formalism
developed in Ref.~\cite{blythe07:_stoch_model}. The Fokker-Plank
equation for the multi-state voter model can be simply obtained as
follows: Let us denote $\conf$ as a generic configuration of the
system (not unique) with $n_i$ voters in state $i$, $\conf = \{ n_1,
n_2, \ldots , n_{S} \}$, with a normalization $\sum_i n_i =N$.  The
probability of finding the system in the configuration $\conf$ at time
$t$, $P(\conf , t)$, evolves in terms of a master equation that is
defined by the transition rates $w (\conf' \to \conf ) $ from the
state $\conf'$ to the state $\conf$. At each time step only one voter
changes its state, consequently we can write a new configuration
$\conf '$ of a transition $\conf \to \conf' $ as $ \conf ' =
\conf_{i+j-} = \{n_1, \ldots, n_j-1, \ldots, n_i+1, \ldots, n_{S} \}$,
being $j$ and $i$ the state of the voter before and after the
transition, respectively. The transition rates $ \conf \to
\conf_{i+j-} $ and $ \conf \to \conf_{i-j+} $ are given by
\begin{equation}
  w(\conf \to \conf_{i+j-} ) =  w(\conf \to \conf_{i-j+} ) =
  \frac{1}{\Delta} \frac{n_i}{N} \frac{n_j}{N},
\end{equation}
where $\Delta = 1/N$ is the natural microscopic time step of the
model, while the transitions rates $ \conf_{i+j-} \to \conf $ and $
\conf_{i-j+} \to \conf $ have the form
\begin{equation}
  w(\conf_{i+j-}  \to \conf) =   \frac{1}{\Delta} \frac{n_i+1}{N}
  \frac{n_j-1}{N}, \quad w(\conf_{i-j+}  \to \conf) =  \frac{1}{\Delta} \frac{n_i-1}{N}
  \frac{n_j+1}{N}. 
\end{equation}
It is now possible to derive the associated master equation. Under the
diffusion approximation \cite{Gardinerbook}, valid for large $N$, we
consider the frequencies of the states $x_i= n_i/N$ and we rescale the
time by a factor $1/N$, so that one time step $t$ measures $N$ updates
of the voters.  A generic configuration is therefore denoted by
$\confx = \{ x_1, x_2, \ldots, x_S \}$, and lies in the standard
simplex $\mathcal{S}_S = \{ \confx \in \mathbb{R} ^S | \sum_i^S x_i =1
\}$.  The set of the $S$ vertices of the simplex, $\mathcal{B}_S = \{
\vec{e}^i \in \mathbb{R} ^S | e^i_j = \delta_{ij}, \; i=1, \ldots, S
\}$ is the absorbing boundary of the dynamics.  We note that the
constraint $\sum_i x_i =1$ reduces the number of independent variables
from $S$ to $S-1$, so we can choose $x_S$ to be dependent on the
others.  Expanding the master equation in terms of $1/N$ we finally
obtain, up to order $N^{-2}$, the final Fokker-Plank equation in
continuous time \cite{blythe07:_stoch_model}
\begin{equation}
  \dd_t P (\confx,t)  =
  \frac{1}{N} \sum_{i=1}^{S-1} \dd_i^2 \left[  x_i (1-x_i) P(\confx,t)
  \right] 
  -\frac{2}{N}  \sum_{j<i}^{S-1} \dd_i \dd_j  \left[ x_i x_j
    P(\confx,t) \right] ,
\label{eq:2}
\end{equation}
where $\dd_i \equiv \partial / \partial x_i$.

The Fokker-Plank equation for the multi-state voter model at
mean-field level does not have a drift term. This implies that
the ensemble average density of each state $\av{x_i}$ is constant in
time. This observation allows to extend the calculation of the exit
probability in the standard voter model to general boundary conditions
in the multi-state case.  Let us define the generalized exit
probability $E_{\mathcal{A}} (\confx)$ as the probability that the
system, starting in some random initial configuration $\confx$, orders
in some configuration $\vec{e}^i \in \mathcal{A}$, being $\mathcal{A}$
an arbitrary subset of the absorbing boundary $\mathcal{B}_S$.  During
the evolution of the system, the average densities are conserved.  Let
us define the quantity $\phi_\mathcal{A} = \sum_{i | \vec{e}^i \in
  \mathcal{A}} x_i$, which is also conserved. In the final consensus
state, $\phi_\mathcal{A}$ will have a value $1$ with probability
$E_{\mathcal{A}}(\confx)$, and a value $0$ with probability
$1-E_{\mathcal{A}}(\confx)$.  We hence obtain the generalized exit
probability
\begin{equation}
  \label{eq:6}
  E_{\mathcal{A}}(\confx) = \phi_\mathcal{A}.
  %= \sum_{i | \vec{e}^i \in    \mathcal{A}} x_i. 
\end{equation}

The consensus time for a given initial condition $\confx$ is given, on
the other hand, by the general equation \cite{Gardinerbook}
\begin{equation}
  \label{eq:T}
  - N  = \sum_{i=1}^{S-1} x_i (1-x_i)  \dd_i ^2 T_N (\confx)
  -2  \sum_{j<i}^{S-1}  x_i x_j \dd_i \dd_j  T_N (\confx ) ,
\end{equation} 
subject to the boundary conditions
\begin{equation}
  \label{eq:7}
  T_N(\confx \in \mathcal{B}_S) =0.
\end{equation}
We can solve in a simple way this equation by noting that the
consensus time $ T_N(\confx)$ has to be symmetric under any exchange
$x_i \leftrightarrow x_j$ with $i, j = 1, \ldots S$. Thus, we can
impose the ansatz form
\begin{equation}
  \label{eq:F}
  T_N(\confx) = \sum_{i=1}^{S} \mathcal{ F } (x_i), 
\end{equation}
where the function $ \mathcal{ F } (x)$ is independent of $S$.
Introducing this ansatz into Eq.~(\ref{eq:T}) we obtain
\begin{equation}
  \label{eq:TGF}
  1 = - \frac{1}{N} \sum_{i=1}^{S} x_i (1-x_i)  \dd_i^2  \mathcal{ F }
  (x_i) = \sum_{i=1}^{S}   \mathcal{ G } (x_i), 
\end{equation}
where we have defined
\begin{equation}
  \label{eq:G}
  \mathcal{ G } (x) = - \frac{1}{N}  x (1-x)  \mathcal{ F '' } (x).
\end{equation}
From Eq.~(\ref{eq:TGF}) and the normalization condition for $x_i$, se
can see that the only possible values of $\mathcal{ G } (x)$ are
$\mathcal{ G } (x) = \mathrm{const} \equiv 1/S$ or $\mathcal{ G } (x)
= x$. Considering now the solution for the $S=2$ case in
Eq.~(\ref{eq:1}), we can see that the correct solution is given by the
second case, which, after integration of Eq.~(\ref{eq:G}),
applying the boundary conditions 
$\mathcal{ F } (1) =\mathcal{ F } (0) = 0$, leads to
\begin{equation}
  \label{eq:8}
  T_N (\confx) = -N \sum_{i=1}^{S} (1-x_i) \ln(1-x_i).
\end{equation}
This solution generalizes the ``entropic'' form corresponding to the
standard voter model, Eq.~(\ref{eq:1}), recovering in a considerably
simpler way the formal result previously obtained in
\cite{Tavare1984119,Cox:1989fk,baxter2007}.
  
From Eq.~(\ref{eq:8}), we can analyze the behavior of the system in
the limit of a large number of initial states. In particular,
considering the homogeneous initial conditions $x_i = 1/S$, we have
\begin{equation}
  \label{eq:9}
  T_N^H(S) = N(S-1) \ln\left(\frac{S}{S-1}\right). \label{e:constime}
\end{equation}
That is, as we could naively expect, the consensus time increases with
the number of states allowed (the system is initially more disordered
 and therefore requires more time to
order), but its growth is very slow and saturates in the
limit $S \gg 1$. In fact, in the worst case scenario in a finite
system, in which $S=N$, we have $T_N^H(S=N) =N(N-1)\ln [N/(N-1)] \to
N$ in the limit of large $N$, being only a factor $1/ \ln(2)
\simeq1.44$ larger that the binary case $S=2$.  This result can be
rationalized considering that, when $S=N$, we are effectively
describing an initial condition in which every agent has a different
state, and the ordering occurs when one of this individuals manages
to impose its state at the population level. It is therefore not
surprising that we recover the $N$ behavior observed in the binary
model when the initial condition consists of a given state of one
species in a population of individuals of the opposite state, and,
crucially, only the runs in which the state of the mutant gets fixated
are considered.  From Eq.~\ref{e:constime} we can obtain the form in
which the saturation  at large $S$ is reached, namely,
\begin{equation}
  \label{eq:15}
  1-\frac{T_N^H(S)}{T_N^H(S=N)} = 1-(S-1)
  \ln\left(\frac{S}{S-1}\right) \simeq \frac{1}{2S} +
  \mathcal{O}(S^{-2}),
\end{equation}
where the last expression is asymptotically valid in the limit of large $S$.

\begin{figure}[t]
  \begin{center}
    \includegraphics*[width=0.7\textwidth]{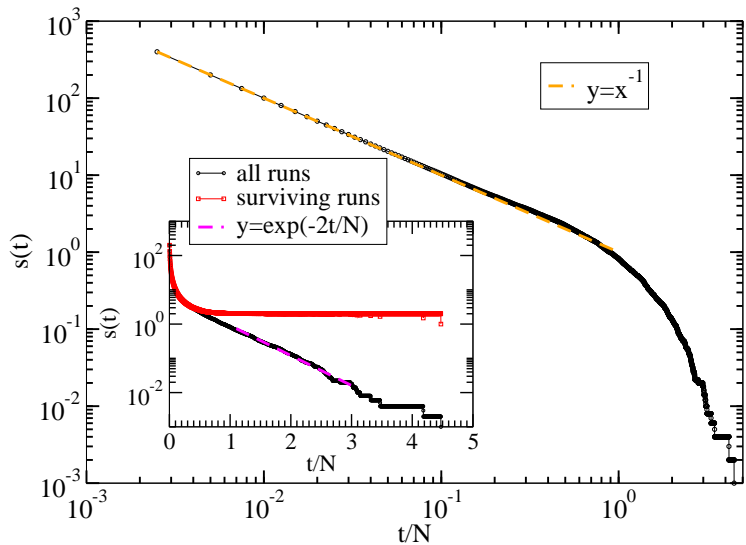}
  \end{center}
  \caption{Number of surviving states $s(t)$ as a function of time in
    the multi-state voter model on a complete graph of size $N=400$,
    starting with homogeneous condition and $s(0)=S=N$ states. We
    compare the result with eq. (\ref{eq:12}). In the inset we show
    $s(t)$ for $t \gg N$, averaged over all runs, compared with
    eq. (\ref{eq:13}).  We observe $s(t) \sim \textrm{const}$ if
    averaged only over surviving runs. }
  \label{fig:S_t}
\end{figure}

Another interesting property of the ordering dynamics of the
multi-state voter model is the number of different states at time $t$,
starting from an initial condition with $S$ states. We define the
number of surviving states as $s(t) = \sum_i \delta_i(t)$, where
$\delta_i(t)=0$ if $x_i(t)=0$, and $\delta_i(t)=1$ otherwise.
Expressions for this quantity have been given in the past in an
implicit form \cite{Tavare1984119,baxter2007}. Here we present a
transparent derivation of its explicit form, based on the form of the
consensus time, Eq.~(\ref{eq:8}).  We start by considering the average
consensus time $\av{T_N(s)}$ for a random initial configuration
$\confx$ with $s$ different states that can be computed by averaging
the consensus time $T_N (\confx)$ over all the initial conditions in
the simplex $\mathcal{S}_s$,
\begin{equation}
  \label{eq:10}
  \av{T_N(s)}= \frac{1}{|\mathcal{S}_s|} \int_{\mathcal{S}_s}d\confx \hspace*{0.1cm}  T(\confx),
\end{equation}
where $|\mathcal{S}_s| = \frac{1}{(s-1)!}$ is the volume of the standard simplex
$\mathcal{S}_s$.
The integral in Eq.~(\ref{eq:10}) can be computed using the variables
$\sigma_n=\sum_i^n x_i$, which respect the constraint $0\leq \sigma_1
\leq \sigma_2 \leq \ldots \leq \sigma_{s-1} \leq \sigma_{s} = 1$, and
noting that
\begin{equation}
  \hspace{-2.5cm} \int_0^{1} \sigma_{s-1} \log( \sigma_{s-1})
  d\sigma_{s-1} \ldots \int_0^{\sigma_{2}}d\sigma_{1} =  \int_0^{1}
  \frac{(\sigma_{s-1})^{s-1}}{(s-2)!} \log( \sigma_{s-1})
  d\sigma_{s-1}  =  \frac{-1}{s^2(s-2)!}. 
\end{equation}
From here it follows that 
\begin{equation}
\label{eq:14}
\av{T_N(s)}=\frac{N(s-1)}{s},
\end{equation}
Now, assuming that the average time to go from $s+1$ to $s$ states is
$\Delta T = \av{T_N(s+1)} -\av{T_N(s)} \simeq Ns^{-2}$, for $s \gg 1$,
we have
\begin{equation}
  \label{eq:11}
\frac{d}{dt} s(t) \simeq \frac{s(t+\Delta T) -s(t)}{\Delta T} = -\frac{1}{\Delta T} \simeq -\frac{s^2}{N}.
\end{equation}
By integrating and inverting this relation we obtain that the number
of surviving states $s(t)$ starting with random initial conditions
with $s(0)=S$ states decays as
\begin{equation}
  \label{eq:12}
s(t) = \left( \frac{t}{N} +\frac{1}{S} \right) ^{-1} \qquad \mathrm{for} \qquad t \ll N,
\end{equation}
expression which is valid far from the ordering time of the system
$\sim N$.  In the case $t \gg N$, we expect $s(t) \sim \mathrm{const}$
in surviving runs; that is, averaging over dynamical realizations that
have not reached consensus at time $t$.  In this case, we will assume
that $s(t)$, averaged over all runs, will decay as the survival
probability \cite{slaninavoter}. Assuming the exponential form derived
in Ref.~\cite{slaninavoter} for the standard voter model, we will
expect to observe
\begin{equation}
  \label{eq:13}
  s(t) \sim \exp(-2t/N) \qquad \mathrm{for} \qquad t \gg N.
\end{equation} 
In Fig. \ref{fig:S_t} we check this prediction by means of numerical
simulations of the multi-state voter model on a complete graph. The
plot shows the behavior of $s(t)$ for homogeneous initial conditions,
which is fully compatible with the analytical predictions in
Eqs.~(\ref{eq:12}) and~(\ref{eq:13}).

\section{Numerical results in finite dimensional lattices}
\label{sec:finite-dimens-latt}

In this section we present and discuss the results of numerical
simulations of the multi-state voter model on lattices of dimension
$d=1$ and $d=2$, comparing them with the analytical results obtained
 at the mean-field level.

\subsection{Consensus time}

We focus in the first place on the behavior of the consensus time
$T_N(\confx)$ as a function of the initial densities of the different
states.  We consider the simplest case $S=3$, parametrizing the
initial configuration as $\confx = \{x_1, x_2, x_3\} \equiv \{x,
\alpha (1- x), (1-\alpha)(1-x) \}$, with $x \in [0,1]$ and $\alpha \in
[0,1]$.  This  parametrization preserves the normalization,
$\sum_i x_i =1$, and has the advantage that, for a given value of
$\alpha$, the whole range of values of $x$ can be explored. 
\begin{figure}[t]
\centerline{\includegraphics*[width=0.55\textwidth]{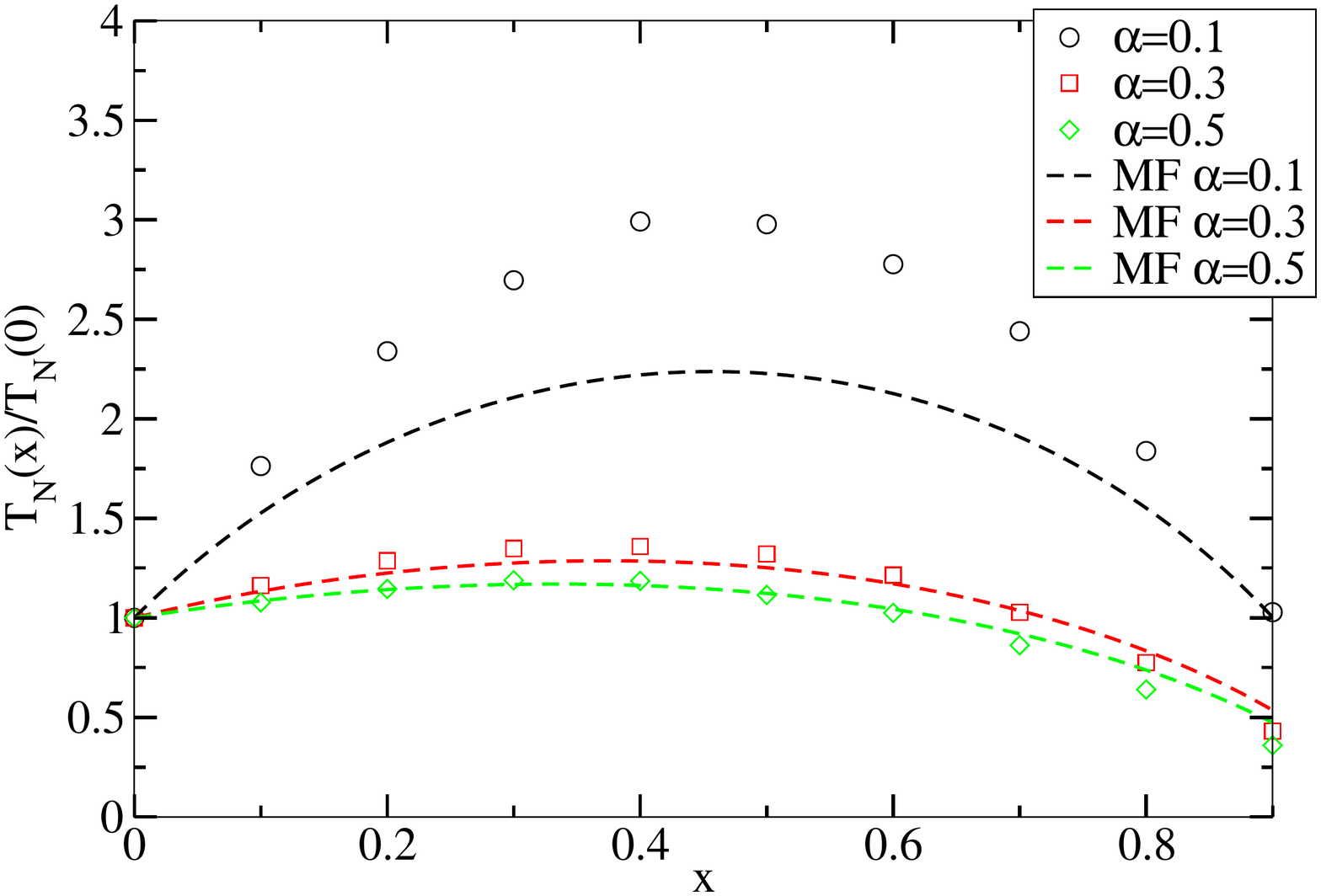}%
\includegraphics*[width=0.55\textwidth]{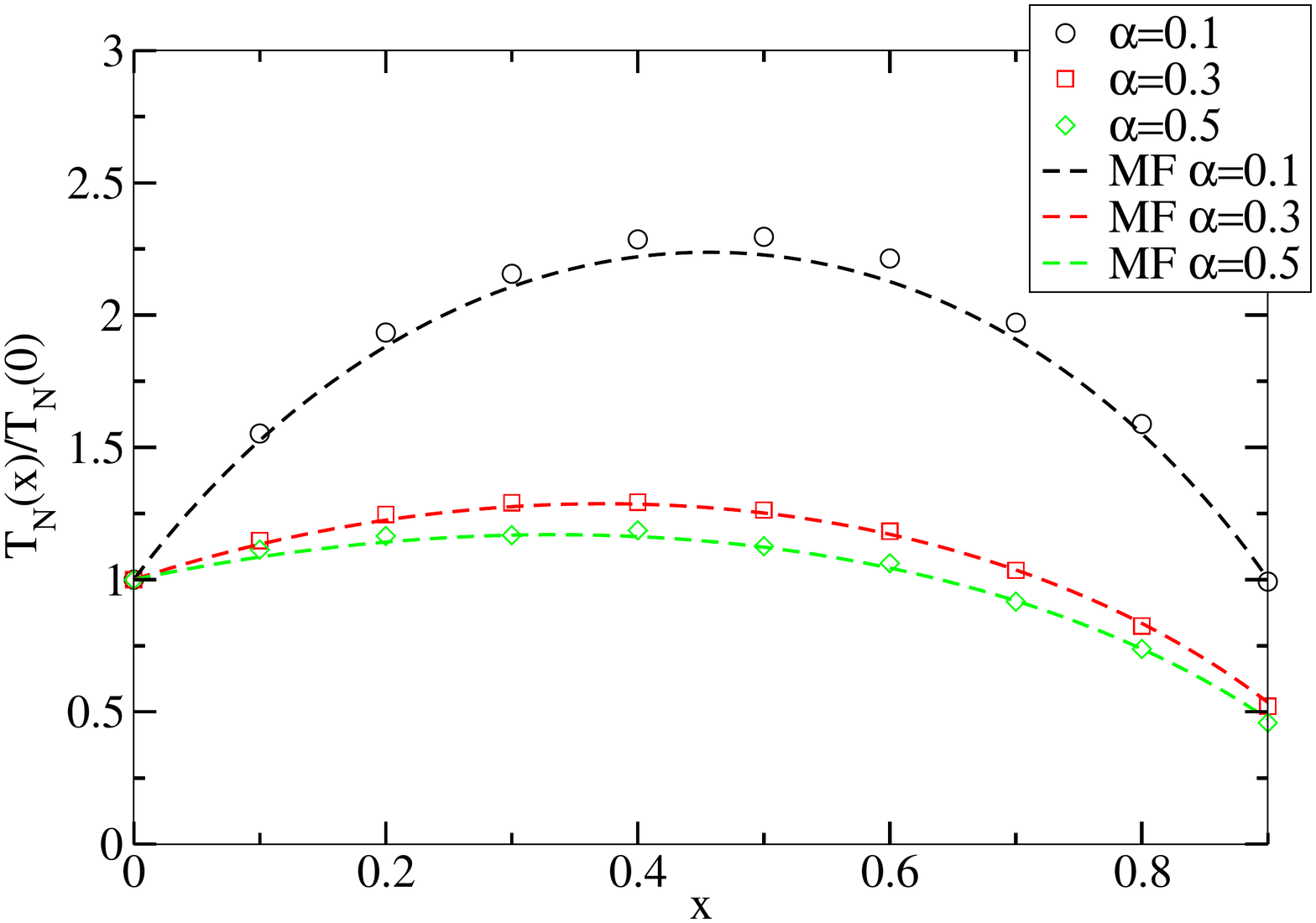}}
\caption{Normalized consensus time $T_N(x)/T_N(0)$ as a function of
  $x$ for the initial configuration $\confx = \{x, \alpha (1- x),
  (1-\alpha)(1-x) \}$ in regular lattices of dimension $d=1$ (left)
  and $d=2$ (right) of size $N=400$ sites, compared with the
  analytical mean-field prediction Eq.~(\ref{eq:8}). }
  \label{fig:alpha1d}
\end{figure}
In Fig.~\ref{fig:alpha1d} we plot the consensus time $T_N(\alpha, x)$
computed in lattices of dimension $d=1$ and $d=2$ as a function of
$x$, and for different fixed values of $\alpha$. In order to get rid
of size-dependent prefactors due to the dimensionality in the
consensus time, we normalized it by its value at $x=0$, which takes
the form $T_N(\alpha) = -N[\alpha \ln(\alpha) +
(1-\alpha)\ln(1-\alpha)]$ at mean-field level. The numerical
simulations for $d=2$ fit quite precisely the theoretical mean-field
prediction of the consensus time dependence on the initial
configuration $\confx$, developed in Sec.~\ref{sec:multi-state-voter},
with only slight deviations for small $\alpha$ and close to $x \sim
0.5$. On the other hand, strong deviations are noticeable in dimension
$d=1$, specially for small values of $\alpha$. This result is in
agreement with the expectation for the standard voter model, in which
the mean-field consensus time Eq.~(\ref{eq:1}) is expected to be exact
only for $d\geq 2$ \cite{1751-8121-43-38-385003}.

\subsection{Effect of the number of states}

\begin{figure}[t]
\begin{center}
\includegraphics*[width=0.7\textwidth]{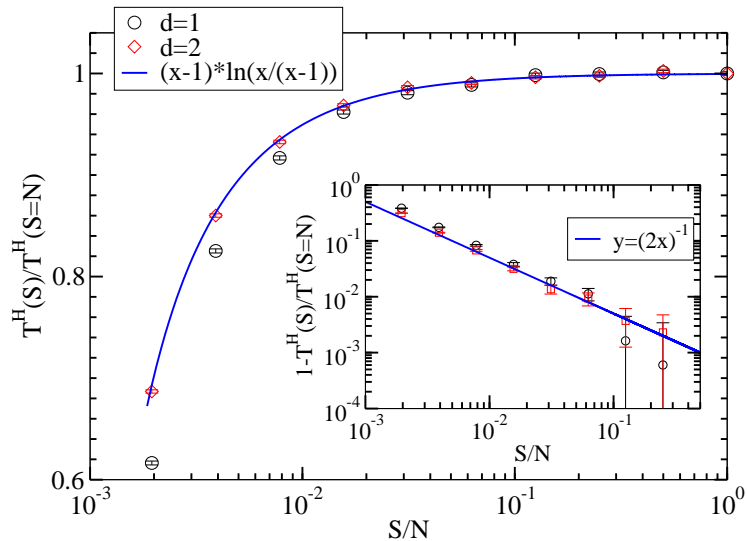}
\end{center}
\caption{Rescaled consensus time $T_N^H(S) / T_N^H(S=N)$ starting from
  homogenous initial conditions, as a function of the number of states
  $S/N$ in dimension $d=1$ and $d=2$, for $N=10^3$, compared with the
  theoretical mean-field prediction Eq.~(\ref{e:constime}). In the
  inset we plot the quantity $1-T_N^H(S) / T_N^H(S=N)$, showing the
  power-law decay with $S$. Error bars represent the standard deviation error 
  on the average of the distribution. Each point is averaged over $10^5$ runs.}
  \label{fig:T_S_hom}
\end{figure}

We have seen that, at the mean field level, and for homogeneous
initial conditions $x_i=1/S$ for $i=1, \ldots, S$, the consensus time
increases with $S$ towards it limit value $T_N^H(S=N)$with a
power-law form, as given by Eq.~(\ref{eq:15}). In
Fig.~\ref{fig:T_S_hom} we plot the rescaled consensus time $T_N^H(S) /
T_N^H(S=N)$ as a function of $S$ for lattices of dimension $d=1$ and
$d=2$ and fixed size $N=10^3$.  From this figure we observe once again
that the $d=2$ behavior is well fitted by the mean-field prediction,
while the $d=1$ case shows deviations for small $S$.  Interestingly,
increasing the number of initial states $S$ reduces the deviation from
the mean-field theory, in a way that the behavior for $S \rightarrow
N$ is very well fitted by Eq.~(\ref{eq:15}).

\subsection{Number of surviving states $s(t)$ }

\begin{figure}[t]
\begin{center}
\includegraphics*[width=0.7\textwidth]{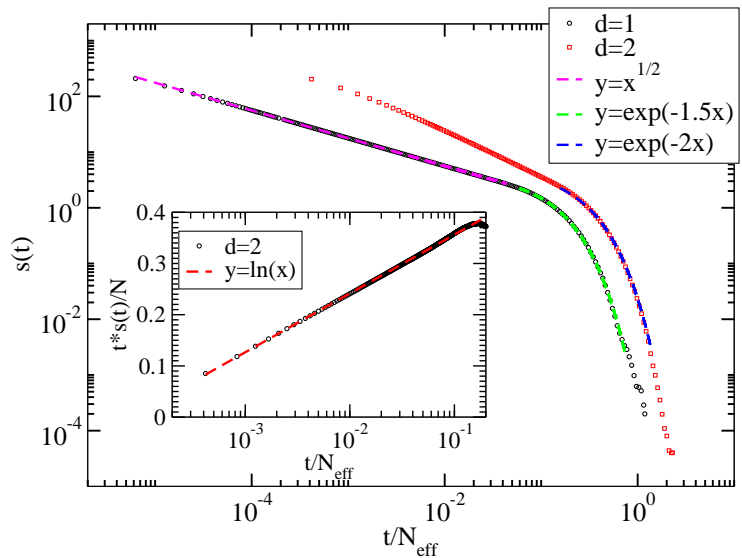}%
\end{center}
\caption{Surviving states $s(t)$ as a function of rescaled time
  $t/N_{\mathrm{eff}}$ for a $d=1$ and $d=2$ lattice of $N=400$ nodes,
  starting with homogeneous condition and $s(0)=S=N$ states.  For $t
  \ll N_\mathrm{eff}$, the number of surviving states decays as $s(t)
  \sim t^{-1/2}$ for $d=1$ and $s(t) \sim t^{-1} \log t$ for $d=2$
  (inset). For large $t$, $s(t) $ decays exponentially in both
  cases. }
  \label{fig:s_t1d2d}
\end{figure}

At the mean field level, the number of surviving
states $s(t)$, starting from maximally heterogeneous conditions $S=N$,
decays as $s(t) \sim S t^{-\beta}$ in the initial time regime, with an
exponent $\beta=1$, crossing over to a exponential decay at large
times.  In Fig. \ref{fig:s_t1d2d} we show the number of surviving
states $s(t)$ as a function of time corresponding to numerical
simulations on $d=1$ and $d=2$ lattices.

From the results of Fig.~\ref{fig:s_t1d2d}, it is clear that the
initial decay of the density of surviving states follows, as expected,
a power-law form. The decay is, however, slower than the mean-field
prediction. In particular, we see that in $d=1$, $s(t) \sim N
t^{-1/2}$, while in $d=2$ numerical data can be fitted to the form
$s(t) \sim N t^{-1} \log t$, corresponding to mean-field behavior with
a logarithmic correction (shown in the inset of
Fig. \ref{fig:s_t1d2d}). On the other hand, we observe that the tail
of the density of surviving states is again exponential; in
particular, we find that in the large time regime we can fit $s(t)
\sim \exp(-1.5 t/N_\mathrm{eff})$ for $d=1$, while $s(t) \sim \exp(-2
t/N_\mathrm{eff})$ for $d=2$.

The origin of the slowing down in the decay of the number of surviving
states in low dimensions can be attributed to the formation of spatial
domains of sites in the same state, which have to annihilate
diffusively in order to reach the consensus state. In this line, the
behavior of the number of surviving states in $d=1$ can be understood
by means of a simple argument: At a given time $t>1$, there will be a
number of surviving states $s(t)$. Assuming that sites with the same
state form clusters, the activity will be driven by the diffusive
fluctuation of the boundaries of those clusters, which will have a
length $\ell \propto t^{1/2}$. The number of different clusters will
thus be $s(t) \propto N / \ell \sim N t^{-1/2}$, recovering the
observed time dependence.

\subsection{Effects of correlated initial configurations}

 \begin{figure}[t]
\begin{center}
\includegraphics*[width=0.7\textwidth]{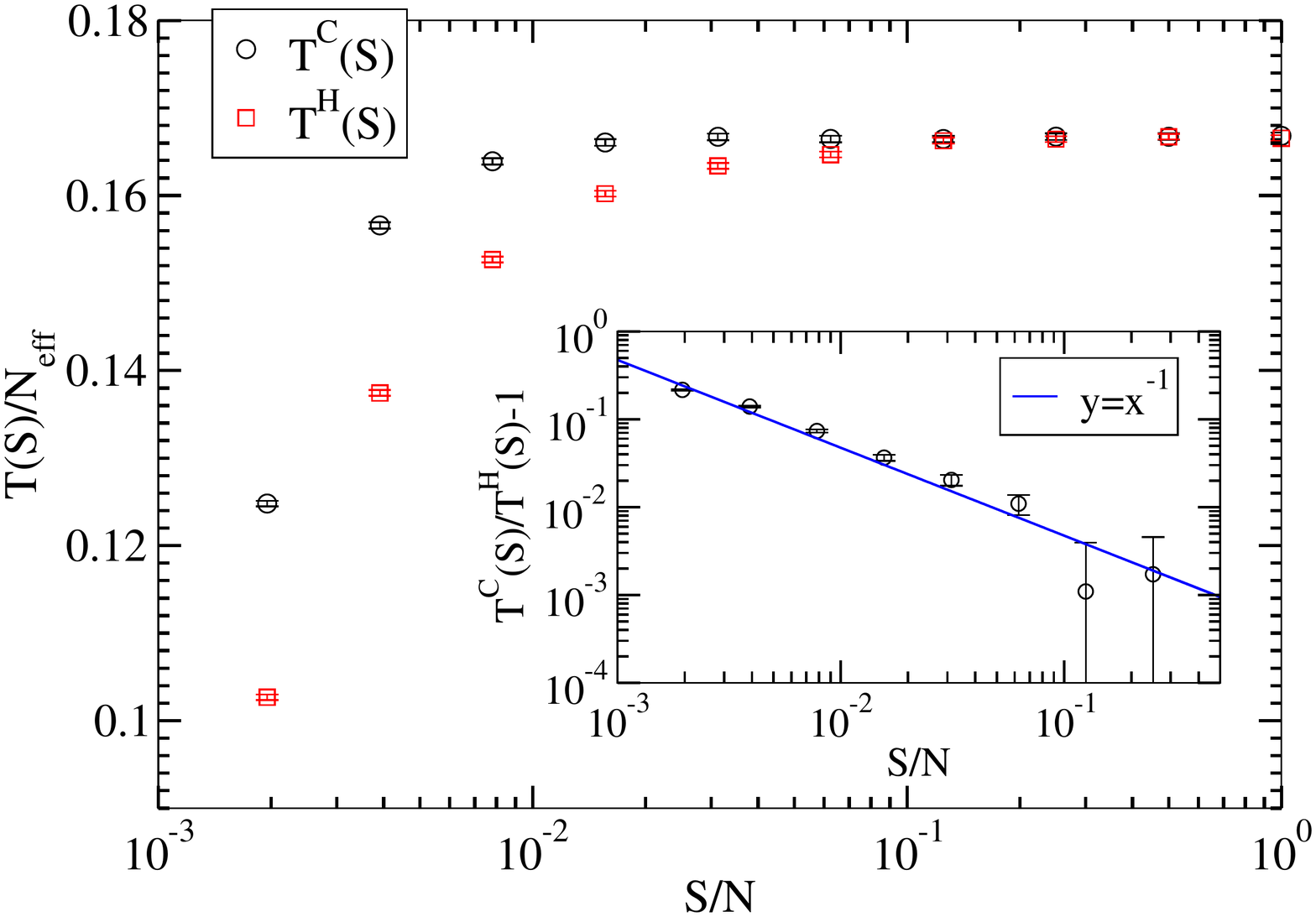}
\end{center}
\caption{Consensus time $T(S)/N_\mathrm{eff}$ as a function of the
  number of initial states $S$ on a $d=1$ lattice with $N=10^3$ nodes
  with a correlated initial configuration made of blocks of voters in
  the same state, of initial length $N/S$, $T^C(S)$, compared with the
  consensus time obtained with random homogeneous initial
  conditions $T^H(S)$w. In the inset we show that the quantity $T^c(S)/T^H(S)-1$
  goes to zero with a power law behavior with exponent -1. Error bars
  are obtained as in Figure \ref{fig:T_S_hom}.  Each point is averaged
  over $10^5$ runs. }
  \label{fig:block}
\end{figure}

Considering the multi-state voter model on a finite lattice allows to
investigate the effects of correlated initial configurations in the
dynamical approach to the consensus state, which should be
particularly important in one-dimensional lattices. We have thus
simulated the multi-state voter model in a $d=1$ lattice, stating from
an initial configuration of $S$ states arranged in $S$ contiguous
blocks on length $N/S$ in a lattice of size $N$. In
Fig.~\ref{fig:block} we plot the consensus time for $T^C_N(S)$ in this
correlated initial conditions as a function of $S$, comparing it with
the consensus time starting with uncorrelated homogeneous initial
conditions, $T^H_N(S)$.  We find that the effect of starting with
correlated initial condition strongly slows down the achievement of
consensus.  As expected, the difference between the consensus time
with correlated and homogeneous initials conditions approaches zero
for $S \rightarrow N$ with a behavior compatible with a power-law form
of exponent $-1$, i.e.
\begin{equation}
\frac{T_N^C(S)}{T_N(S)}-1 \sim \left( \frac{S}{N}-1 \right) ^{-1}
\qquad \mathrm{for }\; S \rightarrow N. 
\end{equation}

\section{Conclusions}
\label{sec:conclusions}

In this paper we have addressed the general scenario of the voter
model in which the number of different states allowed in the model can
be larger than two, and, in the thermodynamic limit, even
unlimited. At the mean-field level, we have presented derivations for
the expression of the exit probability, the consensus time (which
generalizes naturally the `entropic' form observed for the two-states
case), and the density of surviving states as a function of time.  We
have highlighted that in the limit of $S \rightarrow \infty$ the
ordering time is only $1 / \ln 2$ times bigger than in the binary
voter model, and with a simple analytic argument we have found the
decay of the number of surviving states in times.  Finally, we have
studied numerically the behavior of the multi state voter model on
$1-$ and $2-$dimensional lattices, and compared the results with the
binary case.  The consensus time in the $d=2$ case is well predicted
by the mean-field theory, while the uni-dimensional case behaves
differently.  Remarkably, it increases with the number of initial
states with a power-law form as predicted by the mean-field theory,
for both $d=1$ and $d=2$.  We have also addressed the effect of
correlated initial conditions on the consensus time, finding out that
the relevance of this effect decreases with the number of initial
states with a power law behavior.  In summary, our results show that
the number of states is not a trivial parameter in the voter model,
and it affects the overall dynamics in subtle ways.

\section*{Acknowledgments} 
We acknowledge financial support from the Spanish MEC, under project
FIS2010-21781-C02-01, and the Junta de Andaluc\'{i}a, under project
No. P09-FQM4682.. R.P.-S. acknowledges additional support through
ICREA Academia, funded by the Generalitat de Catalunya.

%%%%%%%%%%%%%%%%%%%

\providecommand{\newblock}{}


\begin{thebibliography}{10}
\expandafter\ifx\csname url\endcsname\relax
  \def\url#1{{\tt #1}}\fi
\expandafter\ifx\csname urlprefix\endcsname\relax\def\urlprefix{URL }\fi
\providecommand{\eprint}[2][]{\url{#2}}

\bibitem{Castellano09}
Castellano C, Fortunato S and Loreto V 2009 {\em Rev. Mod. Phys.\/} {\bf 81}
  591--646

\bibitem{Clifford73}
Clifford P and Sudbury A 1973 {\em Biometrika\/} {\bf 60} 581--588

\bibitem{Holley:1975fk}
Holley R~A and Liggett T~M 1975 {\em Annals of Probability\/} {\bf 3} 643--663

\bibitem{liggett99:_stoch_inter}
Liggett T~M 1999 {\em Stochastic interacting particle systems: Contact, Voter,
  and Exclusion processes\/} (New York: Springer-Verlag)

\bibitem{KineticViewRedner}
Krapivsky P, Redner S and {Ben-Naim} E 2010 {\em A Kinetic View of Statistical
  Physics\/} (Cambridge: Cambridge University Press)

\bibitem{1751-8121-43-38-385003}
Blythe R~A 2010 {\em Journal of Physics A: Mathematical and Theoretical\/} {\bf
  43} 385003

\bibitem{PhysRevLett.91.028701}
Mobilia M 2003 {\em Phys. Rev. Lett.\/} {\bf 91} 028701

\bibitem{1742-5468-2007-08-P08029}
Mobilia M, Petersen A and Redner S 2007 {\em Journal of Statistical Mechanics:
  Theory and Experiment\/} {\bf 2007} P08029

\bibitem{0295-5075-77-6-60005}
Dall'Asta L and Castellano C 2007 {\em Europhys. Lett.\/} {\bf 77} 60005

\bibitem{PhysRevLett.101.018701}
Stark H, Tessone C~J and Schweitzer F 2008 {\em Phys. Rev. Lett.\/} {\bf 101}
  018701

\bibitem{Lambiotte08}
Lambiotte R and Redner S 2008 {\em Europhys. Lett.\/} {\bf 82} 18007

\bibitem{durretnonlinear}
Cox J~T and Durret R 1991 Nonlinear voter models {\em Random Walks, Brownian
  Motion, and Interacting Particle Systems\/} ed Durrett R and Kesten H
  (Boston: Birkhauser) pp 189--202

\bibitem{Deoliveira93}
de~Oliveira M, Mendes J and Santos M 1993 {\em J. Phys. A\/} {\bf 26}
  2317--2324

\bibitem{Wupottsmodel}
Wu F~Y 1982 {\em Rev. Mod. Phys.\/} {\bf 54} 235--268

\bibitem{PhysRevE.52.244}
Sire C and Majumdar S~N 1995 {\em Phys. Rev. E\/} {\bf 52} 244--254

\bibitem{1742-5468-2008-05-P05006}
L{\'o}pez F~J, Sanz G and Sobottka M 2008 {\em Journal of Statistical
  Mechanics: Theory and Experiment\/} {\bf 2008} P05006

\bibitem{2012arXiv1201.5198B}
B\"ohme G~A and Gross T 2012 {\em Phys. Rev. E\/} {\bf 85} 066117

\bibitem{hubbell2001unified}
Hubbell S 2001 {\em The Unified Neutral Theory of Biodiversity and
  Biogeography\/} Monographs in Population Biology (Princeton, NJ: Princeton
  University Press)

\bibitem{McKane200467}
McKane A~J, Alonso D and Sol\'e R~V 2004 {\em Theoretical Population Biology\/}
  {\bf 65} 67 -- 73

\bibitem{pigolottieco2005}
Pigolotti S, Flammini A, Marsili M and Maritan A 2005 {\em Proc. Natl. Acad.
  Sci. USA\/} {\bf 102} 15747

\bibitem{0295-5075-85-4-48003}
Volovik D, Mobilia M and Redner S 2009 {\em Europhy. Lett.\/} {\bf 85} 48003

\bibitem{0305-4470-36-3-103}
V\'azquez F, Krapivsky P~L and Redner S 2003 {\em Journal of Physics A:
  Mathematical and General\/} {\bf 36} L61

\bibitem{1367-2630-8-12-308}
Castell{\'o} X, Egu{\'\i}luz V~M and Miguel M~S 2006 {\em New Journal of
  Physics\/} {\bf 8} 308

\bibitem{Tavare1984119}
Tavar\'e S 1984 {\em Theoretical Population Biology\/} {\bf 26} 119--164

\bibitem{Cox:1989fk}
Cox J~T 1989 {\em The Annals of Probability\/} {\bf 17} 1333--1366

\bibitem{baxter2007}
Baxter G~J, Blythe R~A and McKane A~J 2007 {\em Mathematical Biosciences\/}
  {\bf 209} 124--170

\bibitem{blythe07:_stoch_model}
Blythe R~A and McKane A~J 2007 {\em J. Stat. Mech.\/}  P07018

\bibitem{Gardinerbook}
Gardiner C~W 1985 {\em Handbook of stochastic methods\/} 2nd ed (Berlin:
  Springer)

\bibitem{slaninavoter}
{F Slanina} and {H Lavicka} 2003 {\em Eur. Phys. J. B\/} {\bf 35} 279--288

\end{thebibliography}
\end{document}